\documentclass[twocolumn,showpacs,amsmath,amssymb]{revtex4-1}

\usepackage{graphicx} 
\usepackage{bm} 
\usepackage{epsfig}

\begin{document}

\title{Influence of swimming strategy on microorganism separation by asymmetric obstacles}

\author{I. Berdakin$^1$, Y. Jeyaram$^2$, V. V. Moshchalkov$^2$, L. Venken$^3$, S. Dierckx$^3$, S. J. Vanderleyden$^3$, A. V. Silhanek$^4$, C. A. Condat$^1$, and V. I. Marconi$^1$}

\affiliation{$^1$Facultad de Matem\'atica, Astronom{\'i}a y F{\'i}sica, Universidad Nacional de C\'ordoba
     and  IFEG-CONICET, X5000HUA C\'ordoba, Argentina.\\
     $^2$INPAC -- Institute for Nanoscale Physics and Chemistry, Nanoscale Superconductivity and Magnetism Group, K.U.Leuven, Celestijnenlaan 200D, B--3001 Leuven, Belgium.\\
     $^3$Faculty of Bioscience Engineering - Department of Microbial
and Molecular Systems, Katholieke Universiteit Leuven, Belgium.\\
$^4$D\'epartement de Physique, Universit\'e de Li\`ege, B-4000 Sart Tilman, Belgium.\\}

\begin{abstract}
It has been shown that a nanoliter chamber separated by a wall of asymmetric obstacles can lead to an inhomogeneous distribution of self-propelled microorganisms. Although it is well established that this rectification effect arises from the interaction between the swimmers and the non-centrosymmetric pillars, here we demonstrate numerically that its efficiency is strongly dependent on the detailed dynamics of the individual microorganism. In particular,  for the case of  run-and-tumble dynamics, the distribution of run lengths, the rotational diffusion and the partial preservation of run orientation memory through a tumble are important factors when computing the rectification efficiency.  In addition, we optimize the geometrical dimensions of the asymmetric pillars in order to maximize the swimmer concentration and we illustrate how it can be used for sorting by swimming strategy in a long array of parallel obstacles.
\end{abstract}


\maketitle
\section{INTRODUCTION}
In recent years  there has been an increasing interest in the dynamics of  microscopic agents confined to nanoliter-volume
 chambers~\citep{Squires2005, Leung2012}. These agents range from cancer~\citep{Voldman2009, Mahmud2009, Konstan2013},
 fibroblast~\citep{Chang2013} and stem cells~\citep{Peng2011} to  self-propelled bacteria~\citep{Galajda2007, Kim2010},
 human spermatozoa~\citep{Denissenko2012}, and microbots~\citep{Sanchez2011}.  Each  self-propelled agent has a specific
 propulsion mechanism and interacts in its own characteristic way with the confining walls~\citep{Berke2008, Li2009, Gaston2011}.
 A strong motivation for the study of these systems is the possibility to sort out, concentrate and manipulate the movement and
 distribution of the swimmers,  or even to harvest their energy  by   using  suitably designed
 micro-architectures~\citep{Leung2012, Mahmud2009, Konstan2013, Chang2013, Peng2011,  Kim2010, Galajda2007,  Denissenko2012, Gaston2011, Angelani2009, Sokolov2010}.

In a pioneer work, Galajda {\it et al.}~\citep{Galajda2007} experimentally demonstrated the possibility of achieving inhomogeneous
 bacterial concentrations by inducing an asymmetric average bacterial displacement with a micro-fabricated wall of funnel-shaped
 openings. This mechanically driven segregation seems to be threatened at sufficiently high bacterial concentrations when they
 can collectively migrate against the confining barriers by creating a chemoattractant gradient~\citep{Lambert2010}. Very recently  
  a counter-intuitive symmetry breaking  of the strong bacterial concentration was observed  under controlled flow as well, 
 using a micro-fluidic channel with a symmetric single funnel~\citep{Altshuler2013}. 
Clearly, the ability to design these structures is highly dependent on our understanding of the key biophysical
 concepts: the motility mechanisms, the interactions between the agents and the walls, and  the hydrodynamic agent-agent interactions.

Considerable attention has been devoted to motile bacteria propelled by rotary motors and exhibiting run-and-tumble dynamics. During
 the run mode, the flagella rotate counterclockwise and the microorganism moves in a forward, relatively straight direction, whereas
 during the tumble mode, one or more flagella rotate clockwise and the bacterium is reoriented towards a new direction~\citep{Berg2004, Condat2005}.
  In the case of the paradigmatic bacterium {\it Escherichia coli},  the dynamics of its wild-type and two mutants has been previously
 studied by Berg and Brown \citep{Berg1972} using a three-dimensional tracking microscope. These authors found that (i) the run length
 is not a constant, but follows an exponential distribution, (ii) the runs do not consist of strictly straight displacements but the
 cell meanders due to rotational diffusion,  and (iii)  the distribution of changes of direction from the end of one run to the
 beginning of the next has a maximum  at a direction making an acute angle with the trajectory of the precedent run. 

The first numerical attempt to describe the observed rectification of bacterial displacement~\citep{Galajda2007} taking
   into  account some of the above mentioned ingredients  was carried out by Wan and collaborators~\citep{Wan2008}.
 These authors considered  point-like swimming bacteria following  a run-and-tumble  dynamics with a constant motor
 force magnitude and  thermal fluctuations without  taking into account  hydrodynamic effects. This model is able
 to reproduce the most important experimental findings, i.e., the  accumulation of swimmers next to the boundaries and
 the ratchet-like effects of the asymmetric wall of funnels, although  it ignores some important details of the swimmer
 dynamics,  whose consideration, as we will show in this work, leads to a more accomplished and quantitative description
 of the observed phenomena.

 In several previous simulations bacteria were assumed to have fixed run lengths, to emerge from each tumble in a completely
 random direction, and to move all with equal speed. Although these hypotheses capture the essential micro-swimmers mechanisms,
 they are rather artificial or inaccurate when addressing particular species~\citep{Berg1972,  Liu2009, Erglis2007, Stocker2011}.
  The question now arises as to whether the particularities  of the dynamics of each species have an impact on the efficiency of
 a mechanical sorter of the type mentioned above.

In this paper we address this question by  investigating how different swimming strategies influence both  the rectification and
 the  separation of micro-swimmers. In particular, we find that the residual memory of run orientation that remains after a tumble,
 leading to a persistent random walk trajectory, substantially increases the rectification efficiency, thanks to an enhancement of
 the effective diffusion coefficient and a longer dwell time near the walls. Taking this effect into account drastically improves
 the quantitative agreement with previous experiments, in time and magnitude of the rectification.   This refined model allows us
  to determine the optimum geometrical parameters that maximize  rectification, which is
  essential for optimizing future applications in sorting, filtering, or harvesting mixed populations of self-propelled agents.

\section{MODEL}

\begin{figure}[ht]
\begin{center}
\includegraphics[angle=0, width=8.5cm]{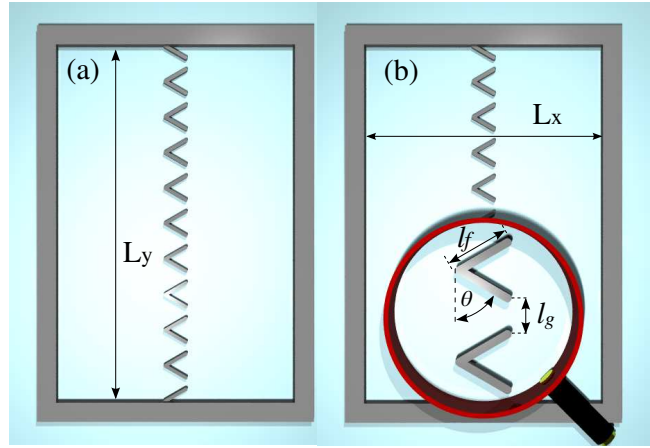}
\caption{(a) Two dimensional reservoir  with a central column of 10 asymmetric funnel-shaped obstacles separating chamber 1 (left) from chamber 2 (right).
   Box size: $L_x \times L_y$.  (b) Details of the ratchet geometry:  $l_f$, $\theta$, and $l_g$, the relevant \textit{control} parameters for a single funnel.}
 \label{fig:parameters}
\end{center}
\end{figure}

We model the dynamics of $N_s$ self-propelled swimmers at low Reynolds numbers, confined to a micro-patterned two-dimensional box
 of size $L_x  \times L_y$,  as schematically shown in Fig.~\ref{fig:parameters}. The swimmer density is $\rho_s = N_s/(L_x \times L_y)$ and
 each individual swimmer, $s_i$,  is represented by a soft disk of radius $r_s$, obeying the following overdamped equation of motion:

\begin{equation}
\gamma \frac{d{\bf r}_i}{dt} = {\bf F}^m_i  + {\bf F}^{sw}_i  +  {\bf F}^s_i , \label{eq:force}
\end{equation}

\noindent where $\gamma$ is the damping constant associated to a given aqueous medium with viscosity $\eta$,
 ${\bf r}_i$ is the position of the swimmer center of mass,   ${\bf F}^m_i$ accounts
 for the internal motor of the swimmer, i.e. the driving force. 
   ${\bf F}^{sw}_i$ is the interaction force between a swimmer and the solid boundaries, and ${\bf F}^s_i$ represents the steric force due to
 the swimmer-swimmer interaction. In the following the forces will be given in units of $\gamma$. We  simulate  {\it diluted} systems with $N_s$ between $1000$ and $3000$ swimmers of the same type inside a rectangular box
 divided by a central line of asymmetric obstacles, Fig.~\ref{fig:parameters}(a), ${\bf F}^s_i$  being the smallest contribution to Eq.~(\ref{eq:force}).   
 The line of obstacles consists of $N_f$ funnels of identical geometry,  $l_g$ being the  gap width between obstacles (opening of the funnel),
 $l_f$ the funnel wall length, and $\theta$ the angle of the aperture (see Fig.~\ref{fig:parameters}(b)).

We assume for the run-and-tumble dynamics that each microorganism  swims  in an {\it almost}
 straight trajectory during a  run, at a  constant speed $| {\bf v}_i |$ for a period of time
 $\tau$ (see scheme in Fig.S1(a) in the Supporting Material~\citep{sup1} ) and that the population of $N_s$
 swimmers has a normal velocity distribution with mean $\bar v$. The swimmer speed $| {\bf v}_i |$ is
 directly proportional to the driving force  magnitude $|{\bf F}^m_i|$. Run lengths, $l_{run,i}=\tau \Delta t |{\bf F}^m_i|$,
 with $\Delta t$ the numerical integration step,    are not  constant in each swimmer trajectory, but have an exponential
 distribution and their direction is changed only after a tumble or due to  the noise affecting its rotational dynamics. 
The stochasticity in the motion is introduced here.  We model the velocity during each run as
 $ {\bf v}_i = | {\bf v}_i | {\hat  {\bf  v}_i}$, with  ${\hat  {\bf v}_i}= \cos(\Phi_i) {\hat  {\bf i}} + \sin ({\Phi_i}) {\hat  {\bf j}}$.
 	The change in direction during a run, 
$\delta \Phi$, in a simulated time interval $\Delta t$, is taken to be  proportional to a gaussian-distributed random
 variable $\nu$: $\delta \Phi = \nu  \sqrt{2 D_r \Delta t }$ \citep{Berg_rw_book}. $D_r$ should be  the coefficient
 of rotational diffusion obtained experimentally for each swimmer species. 
Then we take into account rotational diffusion through the  motor force term.
 In particular to simulate wild-type {\it E. coli} we use $D_r$= 0.1 rad$^2$/sec approximating the
 measured $D_r\sim0.062$ rad$^2$/sec under tracking conditions of $T=305^\circ K$ and $\eta=0.027$ g/cm sec \citep{Berg_rw_book}.   

  The run-and-tumble dynamics was experimentally analyzed by three-dimensional tracking microscopy by Berg and Brown,
 who found that the mean change in direction for wild-type 
{\it E. coli} is $\bar \phi = 68^\circ$ after a tumble, with a standard deviation of $36^\circ$~\citep{Berg1972}.
  In other words, {\it E. coli} keeps some memory of the original direction of motion.  Lovely and Dahlquist~\citep{Lovely1975} studied this effect for cells swimming at a constant speed along a trajectory comprising
 a sequence of exponentially distributed straight runs  of mean duration $\tau$.  They showed that this memory
 increases the effective diffusion coefficient as:

\begin{equation}
D = \frac{v^2 \tau}{d(1 - \langle \cos(\phi) \rangle )} \label{eq:dif}
\end{equation}

\noindent where $\phi$ is the angle between the incoming and the outgoing directions at a tumble
  (See Fig.S1 at~\citep{sup1}), $ \langle ... \rangle$ indicates the mean value,  $d$ is the system dimension, and $v$ is the cell speed.
  If the change in direction from run to run is random, $D=v^2 \tau/d$ \citep{Berg_rw_book}.
  The mean square displacement of the cell is given by the standard expression as a function of time: $\langle \Delta r^2(t) \rangle = 2dD t$.

\begin{center}
\begin{table}
\begin{tabular}{|c|p{1.45cm}|p{1.45cm}|p{1.45cm}|p{1.4cm}|p{1.8cm}|}
\hline
   & Mean direction change, $\bar \phi$~[$ ^\circ$]  & Mean run
duration, $ \tau$[$s$]  & Mean speed, $\bar v$[$\mu m/s $] & $D$ [$\mu m^2/s $] &  Rectification time [$min$]\\
\hline \hline
 $s_1$ &  $33 \pm 15$   & $6.3$    & $20 \pm 4.9$    & $1788$  & $7.1$ \\

 $s_2$ &   $0 \pm 36 $  &  $100$   &  $14.2 \pm 3.4$ & $1293$  & $7.4$  \\

 $s_3$ &  $68 \pm 36$   &  $0.86$  &  $14.2 \pm 3.4$ &  $128$ & $21.4$ \\

 $s_4$ &  $74 \pm 33$   &  $0.42$  &  $14.4 \pm 3.9$ & $48$ & $32.7$  \\

 $s_5$ &  $180 \pm 36$  &  $0.86$  &  $14.2 \pm 3.4$ & $51$ & $33.5$\\

 $s_6$ &  $68 \pm 36$   &  $0.01$  &  $14.2 \pm 3.4$ & $1.6$  & $95.9$ \\

 $s_7$ &  $90$(srw)        &  $0.86$  &  $14.2 \pm 3.4$ & $87$ &  $22.4$ \\
\hline
\end{tabular}

\caption{Motility parameters associated to different swimmers. $s_1$, $s_3$ and $s_4$ are the mutants of {\it E. coli} used
 in~\cite{Berg1972}; $s_2$, $s_5$, $s_6$ and $s_7$ are artificial swimmers (srw = symmetric random walk). The last two columns show
 calculated values. The rectification time was determined as the time needed to get 99 $\%$ of the final rectification $r(\infty) = \alpha /              \beta$ (see Eq.~\ref{eq:r} and Fig.~\ref{fig:comp})} 
\label{tab:swimmers}
\end{table}
\end{center}

In our simulations we take  {\it memory effects} into account, and
 the direction of motion after a tumble, $\phi$,   is chosen from a normal probability
 distribution with the parameters associated to a specific species. In Fig.S1(c) (see~\citep{sup1}) we show
  a sample trajectory   obtained with a mean change in direction $\bar \phi = 60^{\circ}$ with standard deviation $25^{\circ}$.
  The calculations were performed using both mean direction changes
 during runs and changes in direction after tumbles distributed in accordance with observations~\citep{Berg1972}.
 Table I lists  the different swimmers considered in the present work, with $s_1$, $s_3$ and $s_4$ corresponding
 to mutants of {\it E. coli} used in~\cite{Berg1972}, CheC497 (long runs), wild-type AW405 (intermediate runs)
 and Unc602 (short runs) respectively, whereas $s_2$, $s_5$, $s_6$ and  $s_7$ represent idealized swimmers.

It is interesting to compare a real single trajectory under no-confinement with paths obtained from different models.
Fig.S1(b), at~\citep{sup1}, shows a trajectory obtained using the model of Ref.~\citep{Wan2008} with a constant driving force applied
 in a randomly chosen direction after each tumble and a  small thermal force term. Notice that the runs consist of
 apparently straight displacements, which in that model could be made more curvilinear  by adding an unrealistically
 high temperature. Fig.S1(b)   is at odds  with the experimental trajectory shown in Fig.S1(d), obtained in
 Ref.~\citep{Berg1972}, where clearly each run consists of a series of small reorientations around a mean value. 
This observed effect due to rotational diffusion has  been incorporated in our model,  which includes the exponential
 distribution of runs, rotational diffusion and persistence in the motor force term (see Fig.S1(c)). Even though our
 path is in two dimensions,  it still resembles  the particular track called ``a flamenco dancer"; simulated  by E.M.
 Purcell in 1975 and shown in Ref.~\citep{Berg_rw_book}.

 It is worth noting that up to this point, we need only the first term of the r.h.s. of Eq.~\ref{eq:force}, to model a single free swimmer trajectory. 
Let us now describe in detail how the interaction forces and confinement walls are included in our model.

  {\it Interaction swimmer-wall}.
The second term in  the r.h.s. of Eq.~\ref{eq:force}, ${\bf F}^{sw}_i$,   accounts for the  interactions with
  confining walls and obstacles that swimmers could find in their paths. Therefore, it is precisely through
 this term that  the detailed asymmetric  geometry shown in Fig.~\ref{fig:parameters}(b) manifests itself.
  Wall widths (funnels and box) are taken to be similar, $w \sim 2-5 $ $\mu m $, and the disk radius representing
 the swimmer,  $s_i$, is in all cases $r_s$~$=0.5 $ $\mu m$. Each $k$ wall is interacting repulsively with each
 $s_i$  as follows: $ {\bf F}^{sw}_i= F^{sw} ( 1 - r_{ik}/{a} )^{0.1} {\hat {\bf n}_k},$ for  $ r_{ik} < a$
 and ${\bf F}^{sw}_i=0$ otherwise.  Here $a= w/2 + r_s$,  $r_{ik}$  is the distance between the  center
 of mass of $s_i$ and the middle of the k-th wall,  and ${\hat {\bf n}_k}$  is a versor  normal to the wall.
 This force term prevents the crossing of the  swimmers over the walls.
 After a swimmer, $s_i$, hits the k-th wall at an angle $\Theta_i$ formed by the incidence direction $\hat {v_i}$
 and the wall, its trajectory becomes parallel to the wall. The velocity component in this direction is assumed to
 be preserved and its value is given by ${\bf v}_w=| {\bf v}_i | \cos(\Theta_i)  {\hat {\bf t}_k} $ where ${\hat {\bf t}_k}$ is a vector
 tangent to the k-th wall. This is unchanged until either the next tumble or rotational diffusion deflects the direction of motion.
 We always use $F^{sw}/F^m \gg 1$, $F^{sw}$ being
 one order of magnitude larger than $F^m$. In this way we phenomenologically introduce the swimming  and accumulation along walls. 
 Note that our model adds the possibility of tumble-mediated wall detachment, which is strongly dependent on swimming strategy.

 {\it Interaction among swimmers}. 
The last term in the r.h.s. of  Eq.~\ref{eq:force} is the repulsive steric
force,\-  ${\bf F}_i^s=F^s \sum_{j\neq i}^{N_s}  \big(1- |r_{ij}|/2r_s\big) {{\bf \hat r}_{ij}}$, which represents the
 interaction between bacterium $i$ and all other bacteria $j$, each separated by a distance $r_{ij}$ in the direction
 ${\bf \hat r}_{ij}=({\bf  r}_i-{\bf  r}_j)/|r_{ij}|$. This is a short range force approximated by a linearly decreasing force,
 having a  maximum $| {\bf F}^s_i |$ at $r_{ij}=0$ and a zero at  $r_{ij}=r_s$; it is nonzero for $|r_{ij}| < r_s $ and zero otherwise.
 We use $F^{s}/F^m \gg 1$ and $F^s$  of the same order of magnitude as $F^{sw}$. 
Very recent measurements reported in  Ref.~\citep{Drescher2011} for cell-cell and cell-wall
 interactions using {\it E. coli} show that the thermal and intrinsic stochasticity wash out
 the effects of long-range fluid dynamics, implying that physical interactions between bacteria
 are basically determined by steric collisions and near-field lubrication forces. These results
 justify neglecting long-range hydrodynamic forces.
In addition,  we are interested in studying  very diluted systems and rectified motion under confinement,
 situations where we expect only a small contribution of these forces. 

\begin{figure}[ht]
\begin{center}
\includegraphics[angle=0, width=8.8cm]{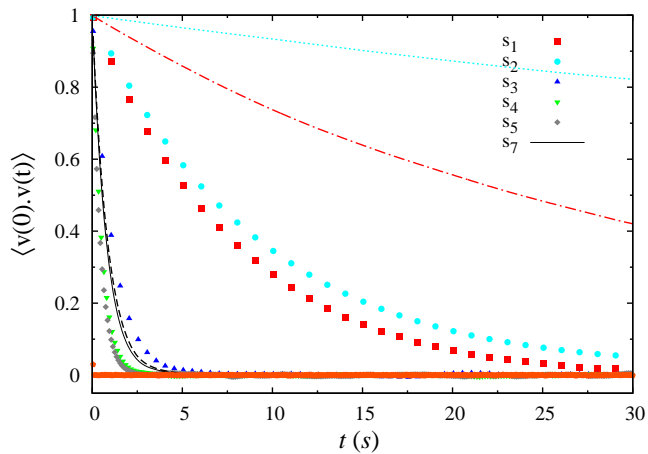}
\caption{(color online) Velocity-velocity correlation function for different self-propelled swimmers.
  Motility parameters for each $s_i$ were taken from Table~\ref{tab:swimmers}. 
 The dot-dashed, dotted, and dashed lines correspond, respectively, to swimmers $s_1$, $s_2$, and $s_7$ when rotational diffusion is not taken into account.}
 \label{fig:corr}
\end{center}
\end{figure}

\section{RESULTS}

\subsection{Swimmers in unbounded environments}

First, we use our  model  to describe the motility properties of run-and-tumble swimmers in the absence of confining surfaces.
 For free swimmers and very diluted systems, our results will be dominated by the first force term  ${\bf F}^m$ in 
 Eq.~\ref{eq:force}. There are two independent processes that degrade the orientational correlation: tumbling and rotational diffusion.
Their combined effect is reflected in the velocity correlation function, $C_{vv}(t)=\langle {\bf v}(t) . {\bf v}(0) \rangle$, shown
 in Fig.~\ref{fig:corr} for the swimmers listed in Table~\ref{tab:swimmers}. 
Note that most of the group of seven swimmers $s_i$, used for comparing different swimming strategies, have the same mean
speed and that the  parameters used for comparison are parameters  not often included  in previous  numerical simulations.
 The comparative  parameters used here characterize each species: the mean change in direction  after a tumble,
 ranging from $\bar \phi = 0^\circ$(high persistence) to $\bar \phi = 180^\circ$ (run and reverse), and the  exponentially distributed
 run lengths, which are proportional to the mean duration of the runs,  ranging from very short ($\tau=0.01$ s) to long ($\tau=100$ s) runs.

 Fig.~\ref{fig:corr} shows clearly that longer runs and a stronger persistence (corresponding to a smaller average directional change
 $\bar \phi$ at each tumble) yield longer correlation times. The correlation for the run-and-reverse case, $s_5$, decays faster than that
 for $s_3$, which has the same run duration but less built-in persistence. Note also that $s_7$ 
 decays faster than two real examples of {\it E. coli} mutants, $s_1$ and $s_3$, because the direction of $s_7$
 is completely randomized at each tumble.
  A simple exponential fit (not shown) to the $s_7$ curve yields a decorrelation time of $0.8$ sec, barely shorter
 than the mean run duration of $0.86$ sec, indicating the relatively weak influence of rotational diffusion. Further
 evidence for the lack of relevance of rotational diffusion for short-run swimmers is the very small difference between
 the results for $s_7$ with (solid line) and without (dashed line) rotational diffusion. The curves for $s_1$ and $s_2$, on the other hand, 
  exhibit long correlations with decorrelation times equal  to $\tau_{vv,s1}=7.9$ $s$ and $\tau_{vv,s2}=8.9$ $s$ respectively.
  Once rotational diffusion is removed, both $s_1$ (dot-dashed line) and $s_2$ (dotted line) decay much more slowly,
 in $\tau_{vv,s1}=43.2$ $s$ and $\tau_{vv,s2}=138.7$ $s$ respectively,  and the difference between their correlations is
 strongly enhanced, showing that, for long-run swimmers, rotational diffusion is the fastest path to memory loss.
 From the precedent discussion it is clear that self-propelled swimmers following run-and-tumble dynamics can
 be classified into two categories: in the first group, typified by $s_1$ and $s_2$, the loss of correlation is
 mostly due to rotational diffusion; in the second group, $s_3$ to $s_7$, decorrelation is controlled by the specific swimming strategy.

\begin{figure}
\begin{center}
\includegraphics[angle=0, width=8.8cm]{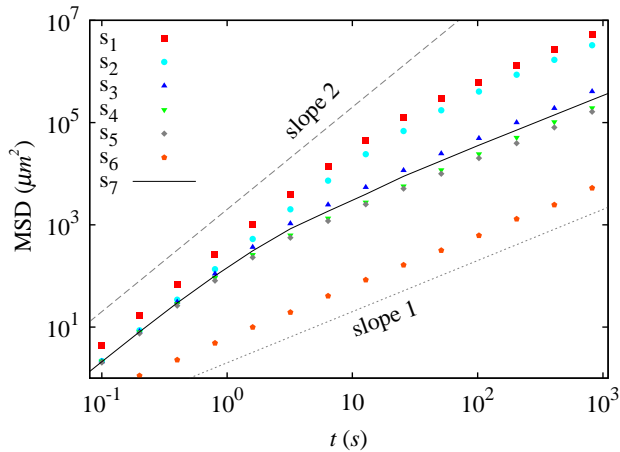}
\caption{(color online) Mean square displacement  for the swimmers  of Table~\ref{tab:swimmers}.  In all cases except for the
 Brownian-like swimmer, $s_6$, the time evolution of the MSD changes from a slope $2$ during ballistic motion (dashed line)
 to slope $1$ during the diffusive regime (dotted line).
 Diffusion increases for larger $\tau's$ and decreases for larger $ \bar \phi 's$. 
 The self-diffusion coefficients obtained for the different swimmers are given in Table~\ref{tab:swimmers}. $D_3$ is
in good agreement with experiments~\citep{Berg2004}. }
\label{fig:dif}
\end{center}
\end{figure}

In Fig.~\ref{fig:dif} we present the mean square displacement (MSD) as a function of time for all $s_i$ considered, taking
 averages over a population of $N_s=1000$ swimmers. In all cases, except for the short-run swimmer $s_6$ (orange pentagons),
 for which the change occurs at very short times, the slope of the time evolution of the MSD is seen to change from $2$
 (dashed line), during ballistic motion, to $1$ in the diffusive regime (dotted line). This transition
 occurs later for those swimmers with longer runs and smaller average directional changes. For swimmers $s_3$ to $s_4$,
 the diffusion coefficient is approximately given by its value in the absence of rotational diffusion (see Eq.~\ref{eq:dif}).
 Note, for instance, the strong reduction in the translational diffusion coefficient resulting from changing the mean angle
 from $68^\circ$ , $s_3$, to $180^\circ$ , $s_5$. The results for $s_7$ swimmers, for which $\bar \phi = 90 ^\circ$ and whose $\tau$ is equal to that of $s_3$, are intermediate between those for $s_3$ and $s_5$,  in reasonable agreement with Eq.~\ref{eq:dif}. The diffusion coefficients
 for $s_1$ and $s_2$ are controlled by rotational diffusion instead. As a consequence, a hundredfold increment in the run
 duration from $s_3$ to $s_2$ generates only a tenfold increment in $D$. The simulation values of $D_3$=128.3 $\mu$m$^2$/s
 and $D_4$=47.94 $\mu$m$^2$/s are also in good agreement with the values obtained by applying Eq.~\ref{eq:dif} to the experimental
 results reported in Ref.~\cite{Berg1972}. These results show
clearly the importance of taking into account the detailed motility properties (specific swimming strategies) for making accurate
 predictions, for instance,  of swimmer separation in mixed species systems.  Both precise numerical calculations and
  accurate experimental data characterizing the swimmer dynamics are therefore of paramount importance.  Better motility statistical
 characterization with improved techniques~\cite{Poon2011} is also needed for each species.
 
\subsection{Swimmers in confined environments}

{\it Wall accumulation.} The specific parameters characterizing each swimmer motility have a strong influence on the time that swimmers
 spend near the walls, a fact crucial to characterize and to use in applications aiming to control and direct swimmer motion with
 micro-geometrical confinements.
 The near-wall accumulation effect is illustrated in Fig.~\ref{fig:wall}.  These 2d simulations were performed for all $s_i$ under
 the same conditions: a population of swimmers, $N_s$, is introduced in  a square  obstacle-free box, i.e  without the funnels shown
 in Fig.~\ref{fig:parameters}. 
  The  density  distribution versus distance to the walls is measured, averaging over the four walls and time.     
Figure~\ref{fig:wall} shows that a larger $\tau$ leads to an increment  in
 the time of permanence close to the walls, which, in turn, enhances the swimmer accumulation near the walls. This explains the
  large accumulation of swimmers $s_1$ and $s_2$ as compared with that of other  swimmers. 
 In particular, if we compare $s_1$, a real non-tumbling {\it E. coli}, with $s_3$,  a tumbling mutant strain,  we observe that $s_1$
has a near-wall probability density more than twice as large as   $s_3$, in good agreement with recent measurements reported in Ref.~{\citep{Gaston2011}.  
 The role played by the mean angle $\bar \phi$ turned after tumbles can also be analyzed following the progressive decrease in the accumulation of
 swimmers $s_3$, $s_4$ and $s_5$, for which the mean speed, $\bar v$, and mean run duration, $\tau$, are kept on the same order of magnitude (see table I).
  This can be understood as a competition between the diverting effect caused by tumbles and the directional persistence due
 to the small rotation angle around the mean direction of motion. The accumulation for swimmers $s_3$ is consistent with experimental results
 shown in Ref.~\citep{Berke2008, Li2009} for {\it E. coli.} 

\begin{figure}
\begin{center}
\includegraphics[angle=0, width=8cm]{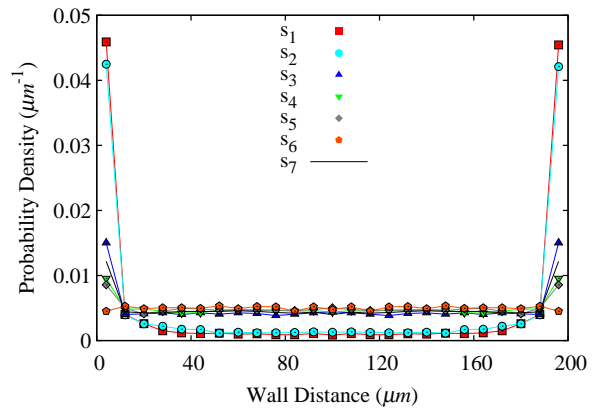}
\caption{ (color online) Accumulation of seven different swimmers (see Table~\ref{tab:swimmers}) near the walls of a  2d container.
 Swimmers were confined to a $ 200 \times 200 $ $\mu m^2 $  box.
} \label{fig:wall}
\end{center}
\end{figure}

\begin{figure*}[bt!]
\begin{center}
\includegraphics[angle=0, width=16cm]{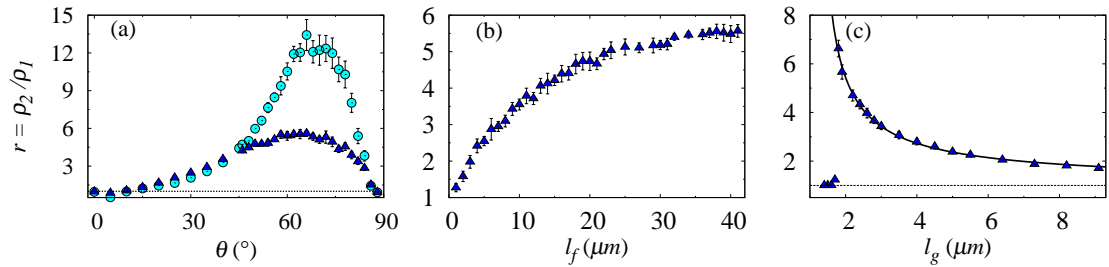}
\caption{ (color online) Rectification efficiency, $r = \rho_2/\rho_1$, vs geometrical parameters:
(a) Effect of changing the funnel angle, $\theta$, for swimmers $s_2$ (upper curve, circles) and $s_3$ (lower curve, triangles). (b)
 Funnel length dependence: $r$ vs $l_f$ for $s_3$.  (c) Gap width dependence, $r$ vs $l_g$ for $s_3$. After a threshold equal to swimmer
 diameter plus  wall width, rectification decreases with  a wider funnel gap as $ 1+a/(l_g-b)^\gamma $ with $\gamma = 0.783 $ for $s_3$.
In (a): 
  $ N_f = 10$, $l_f = 15$ $\mu m $, $l_g = 2$ $\mu m$. While changing one of the parameters the box width is scaled to keep the other two parameters
 unchanged. In (b) and (c), $\theta = 65^\circ$.
 The number of swimmers in the box is also scaled to keep a constant
area fraction  $A_s/A_{box}= 0.0785 $. 
  }
\label{fig:geom}
\end{center}
\end{figure*}

{\it Single  wall of asymmetric funnels between two chambers.}
We next  investigate the dependence of rectification efficiency on both, {\it geometrical and dynamical parameters}. We use a box  divided
 into  two equal chambers, 1 (left) and 2 (right),   by a wall of asymmetric micro-designed funnels,  as originally used in  Ref.~\citep{Galajda2007}
 and illustrated in Fig.~\ref{fig:parameters}. In Ref.~\citep{Galajda2007} it is reported that the easy direction of motion for this kind of
 geometrical rectifier is from chamber 1 to chamber 2.  Knowing  that bacteria  swim along walls it is possible  to gain control on their
 motion  by optimizing the funnel geometry according to  the swimming strategy. Our aim is to design the most efficient rectifier to
 concentrate different self-propelled swimmers in chamber 2,    and later use this knowledge to separate efficiently  mixed populations.  
Let us start by showing the final results (Fig.~\ref{fig:geom}) of a  rectification measurement, defined as the ratio of densities in each
 chamber, $r(t)=\rho_2(t)/\rho_1(t)$, after reaching their steady state. In simulations this quantity is obtained after inoculating the
 swimmers with the same concentration in both chambers $\rho_2 (t=0) = \rho_1(t=0)$. The initial state of the system is prepared as
 a population uniformly distributed over the box and then $r(t)$  is measured. 
Each point in Fig.~\ref{fig:geom}
corresponding to a different geometrical parameter, $\theta$, $l_f$, $l_g$, has been obtained taking special care to
 keep all other parameters fixed, including the swimmer density, by scaling the  box width.
 We studied these dependences for all swimmers $s_i$ in Table~$\ref{tab:swimmers}$.
The dependence of rectification on the angle of the funnels, $\theta$,  both for an artificial
swimmer $s_2$ (upper curve) and the wild type {\it E. coli} $s_3$ (lower curve), presents a
 well-defined peak  around $65^\circ$ independently of  the swimming strategies (not all shown
 in Fig.~\ref{fig:geom}(a)), but with   markedly different maximum values.  Meanwhile the directing
 process becomes more efficient for longer funnel walls until saturating for walls longer than
 $40 $ $\mu m$  for the real swimmer $s_3$, as shown in  Fig.~\ref{fig:geom}(b). Two points need
 to be highlighted in Fig.~\ref{fig:geom}(c). First, swimmers begin to pass through the funnels
 for a gap of size $l_g = 1.7 $ $\mu m$, which is slightly smaller than the sum of swimmer diameter
 and funnel wall width, $w=2 $ $\mu m$, due to the softness of the interactions. Second, once $l_g$
 is above this threshold, the rectification decreases as an inverse power law with exponent $\gamma=0.783$,
 which is slightly smaller than $\gamma=1$ as proposed previously with a different model~\citep{Wan2008}.
 These two results together imply an optimum value of $l_g = 1.8 $ $\mu m$ for our system, always in the
 range of the cell size.  Regarding strictly the ``single funnel geometry'',  it is important to emphasize
 that the angular dependences for designing an optimal rectifier device with a single wall of funnels
 are shared by all swimming strategies studied.  
The optimal values of the main parameters  ($\theta$, $l_f$ or $l_g$) agree with those reported in previous
 studies that consider a single strategy to swim~\citep{Wan2008} but there are important changes in the
 magnitude of rectification and on the time required to achieve it (see Fig.~\ref{fig:comp}). These
 results are, for $s_3$, in good quantitative agreement with experiments~\citep{Galajda2007}. 
The geometrical results obtained   show that neither small angle apertures, corresponding to
 obstacles nearly perpendicular to the rectification direction nor  angles close to  $90^\circ$,
 representing obstacles closely parallel to the current direction  are profitable (Fig.~\ref{fig:geom}(a)).
 Regarding the length of the funnels,  a short $l_f$ is expected  not to be optimal for rectifying because
 the bacterium cannot take  advantage  of the near-to-the-wall swimming  for directing its motion, but  a long $l_f$
 implies a longer time of trapping by the walls in the desired direction until reaching a constant $r$ value. Naturally,
 the optimal opening width, $l_g$, will be  close to the cell size, preventing reentrance or back current swimming,
 which is easier  in wider gap designs.  Small gaps do not exhibit  jamming drawbacks due to the very low concentrations
 considered in this work, for which the probability of clogging is negligible. In conclusion, the best single funnel geometry
 could be predicted in advance with accurate simulations that introduce the specific swimmer motility parameters from good
 measurements. This interdisciplinary approach is  very useful for experimentalists during the initial stage  of designing
 the masks for the lithography process.  For example, if the swimmer is $s_3$, we are sure that the best geometrical
 parameters to be chosen  are $\theta \sim 65^\circ$, $l_f \sim 30 $ $\mu m$ and $l_g \sim$ the cell size, for an expected efficiency of $r \sim 5$.

\begin{figure}
\begin{center}
\includegraphics[angle=0, width=8.5cm]{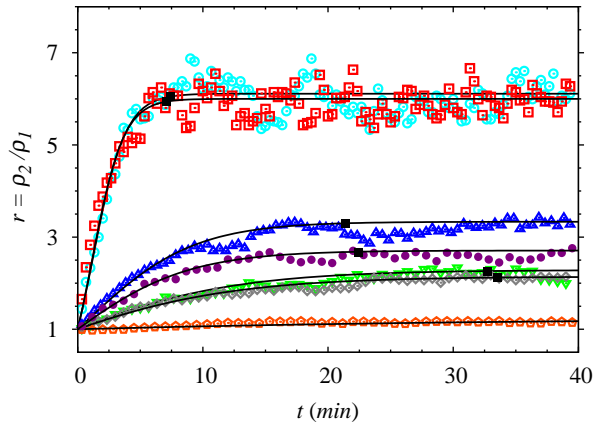}
 \caption{(color online) Rectification  vs time for  swimmers from Table~\ref{tab:swimmers}: $s_1$(red square),
 $s_2$(light blue circle), $s_3$(blue up triangle), $s_4$(green down triangle), $s_5$(grey rhombus), $s_6$(orange pentagon), $s_7$ (filled purple circle).
   Parameters: $N_f = 13$, $l_f = 27 $ $\mu m$, $l_g = 3.8 $ $\mu m$, $\theta = 60^\circ$  and  $L_x=L_y=400 $ $\mu m$. The curve for
 wild type {\it E. coli}, $s_3$, is in excellent agreement with the experimentally determined time to reach the steady state and the final rectification value found in \citep{Galajda2007}.
 Data are fitted by Eq.~\ref{eq:r}. The end of the transitory  period (black filled squares) is calculated as the time needed
 to get to $99 \%$ of the final rectification $r(\infty)=\alpha/\beta$.} \label{fig:comp}
\end{center}
\end{figure}
\begin{figure}
\begin{center}
\includegraphics[angle=0, width=8.8cm]{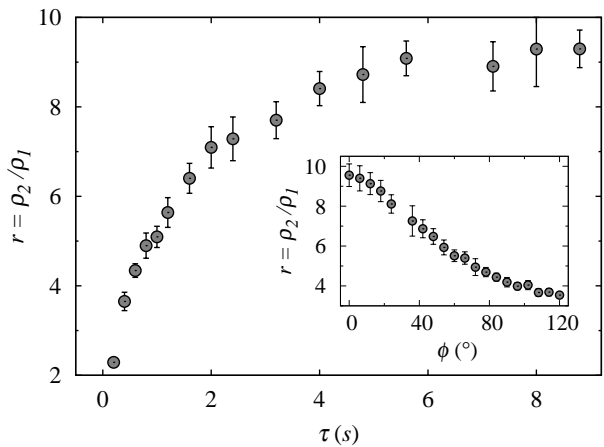}
\caption{Dynamical parameters dependences. Rectification, $r$, vs $\tau$ with persistence $\bar \phi$ = $60^\circ$. Inset: $r$ vs $\bar \phi$ for
 $\tau$=1.0 s. Geometrical parameters used:  $N_f$ = 10, $l_f$ = 15 $\mu m$,~{$l_g$ = 2 $\mu m$},
 $\theta$ = $60^\circ$ and $ L_x$ = $L_y$ = 136 $\mu m$.}
\label{fig:runangle}
\end{center}
\end{figure}
In Fig.~\ref{fig:comp}  we show how the swimming strategies of different swimmers influence the time dependence
 of the  efficiency of the mechanical rectifier. Swimmers with long runs, such as $s_1$ and $s_2$, achieve almost
 twice the rectification  of $s_3$ \citep{Galajda2007} and very quickly, in 10 minutes. Note that even though
 $s_1$ has higher persistence and larger mean speed than $s_2$, both are rectified similarly, showing that the
 considerably longer runs of $s_2$ compensate for its slower motion  and lower persistence. 
A similar analysis could be done for $s_4$ and $s_5$. Since both have the same mean speed and
 $s_5$ has twice the run duration of $s_4$, a quicker rectification for $s_5$ could be expected.
 However, this advantage is lost because the $s_5$ strategy involves reversals in the direction
 of motion. The results for $s_6$, which represents a typical Brownian swimmer with very
 short runs, show that it is not possible to obtain  directed motion if the mean run duration is negligible:
 these swimmers cannot profit from swimming along the walls  as in the previous cases. Finally the rectification time for the $s_7$ swimmer is very similar to the one of the wild type strain because they have the same run time duration and only a $22^\circ$ difference in the mean persistence.

 A simple model for the rectification by a single barrier can be obtained by considering two identical chambers
connected by an asymmetric channel~\citep{Galajda2007}. If the initial concentrations in both chambers are the same, and bacteria cross
from chamber $1$ ($2$) to chamber $2$ ($1$) at a rate  $\alpha$ ($\beta$), it is easy to see that the rectification evolves 
in time as,

\begin{equation}
r(t) = \frac{\rho_2(t)}{\rho_1(t)} = \frac{2 \alpha + (\beta - \alpha) e^{- (\alpha + \beta)t}}{2 \beta - (\beta - \alpha) e^{- (\alpha + \beta)t}} \label{eq:r}
\end{equation}

In all cases, a very good fit is obtained using Eq.~\ref{eq:r}. The fitting parameters are presented in Table S1 at~\citep{sup2}.

In general, there is a one-to-one correlation between how much swimmers get rectified by the funnels and how much
 they diffuse (Fig.~\ref{fig:dif}) or get trapped near walls (Fig.~\ref{fig:wall}). The same is also true for the
 time elapsed to reach the steady state (see the values in Table I). This time was determined as the necessary 
time for the system to present 99\% of the final rectification.   The specific dependence of rectification on
 the strategy of motion was further investigated due to its importance, varying the mean run time, $\tau$,
 and persistence, $\bar \phi$, continuously, as it can be seen in Fig.~\ref{fig:runangle}. In coincidence with
 Fig.~\ref{fig:comp}, these results demonstrate that swimmers with longer runs can be rectified more efficiently.
 At very short runs,  we observe a rapid growth of the final rectification,  which goes from $r=1$  to $r=5$ when
 $\tau$ goes from zero to 2 seconds. The value of $r$ saturates for $\tau $ larger than about $5$ $s$.  This last
 result agrees with Fig.~\ref{fig:comp}, where we see that two swimmers with remarkably  different run durations, $s_1$
  with $\tau=6.3$ $s$ and $s_2$ with $\tau=100$ $s$, show the same $r$ value at long times, even though
 $s_1$ has higher persistence, $\bar \phi = 33^\circ$.

The inset shows that $r (\bar \phi)$ follows the opposite trend, reducing the rectification, as $\bar \phi$ increases. 
For example, compare $s_1$ and $s_3$, both {\it E. coli} mutants. Swimmer $s_1$, with $\bar \phi=33^\circ$, is twice more
 efficient than $s_3$, which has a much lower memory of direction after tumbles, $\bar \phi=68^\circ$.
  The most inefficiently rectified swimmers are those following true random walks paths, as $s_7$ with
 $\bar \phi = 90^\circ$, and the most efficient  micro-swimmers  are those with the  highest memory of
 the previous direction of motion. 
 
 In short, it is important to note here that after optimizing the geometrical parameters, \emph{ the dynamical parameters have a strong influence
 on the efficiency of the rectifiers}, as it is evidenced by the  results shown in Fig.~\ref{fig:comp}  and Fig.~\ref{fig:runangle}.
 Then we emphasize again that for designing optimal devices a previous and accurate motility characterization of each species is
 needed~\citep{Poon2011} with better details in runs distribution and persistence. It is important to characterize
 each mutant motility not only in free swimming but close to surfaces~\citep{Gaston2011} and inside constrictions/channels with dimensions comparable to the cell sizes. 

{\it Array of parallel funnel walls.} 
For extra corroboration of our model, using a geometry consistent of a large  box with five equally distanced and parallel columns of $N_f$=10 asymmetric
 funnels (six chambers from left to right, ranging from 1 to 6)  we inoculated in one extreme, in chamber 1, 
a homogeneous population of $s_3$ swimmers. The simulations reproduce  
 the  exponential increase of the swimmer concentration per consecutive chamber as it was found experimentally~\citep{Galajda2007} (see Fig. S2 in the
 Supporting Material~\citep{sup1}). This exponential distribution of swimmers along the box is due to  the geometric progression of the rectification provided by
 the  successive walls. We studied the dependence on the number of funnels in each wall, $N_f$,
 concluding that it is very weak.
  Then, as we aimed to separate mixed swimmers populations using their swimming strategies, we simulated larger and narrower boxes instead, designs
 closer to tubes or channels geometries, using the most efficient geometrical parameters for each opening as obtained previously.   We  simulated
 a mixture of swimmers in  a  $630 \times 64 $ $\mu m^2$ confinement box separated in 21 identical chambers divided by 20 parallel columns, now with
 only 2 funnels each. For the initial state, the swimmers $s_1$, $s_3$ and $s_4$, {\it E. coli} mutants used in Ref.~\citep{Berg1972}, were inoculated
 in equal proportions  in the first chamber.  
  From the precedent analysis of the effects of a single-wall array, we
 expect that the various populations will begin to move away from the  inoculation chamber,
 even without  chemical attraction or nutrient or temperature gradients, simply by physical guidance
 and motility. Without the funnel array, the expected final state at long time, 
  determined by  active diffusion as in section III.A, 
 is a homogenous distribution of the mixed swimmers over the whole system.  But we observe that the
 micro-swimmer  mixture moves along the array in the easy ratchet direction with the typical rectification
 parameters associated to their respective dynamical parameters, until arriving at  the opposite end of
 the box,  the fastest arriving before 2 min (see second snapshot in Fig.~\ref{fig:snap}). In the
 first panel of Fig.~\ref{fig:snap} we show the initial state of the system, with all swimmers accumulated
 at the left end of chamber 1, from where  they start moving through the different chambers of
 the box with a mean speed that is  strongly dependent on the strategy of motion of the swimmers.

\begin{figure}
\begin{center}
\includegraphics[angle=0, width=8.8cm]{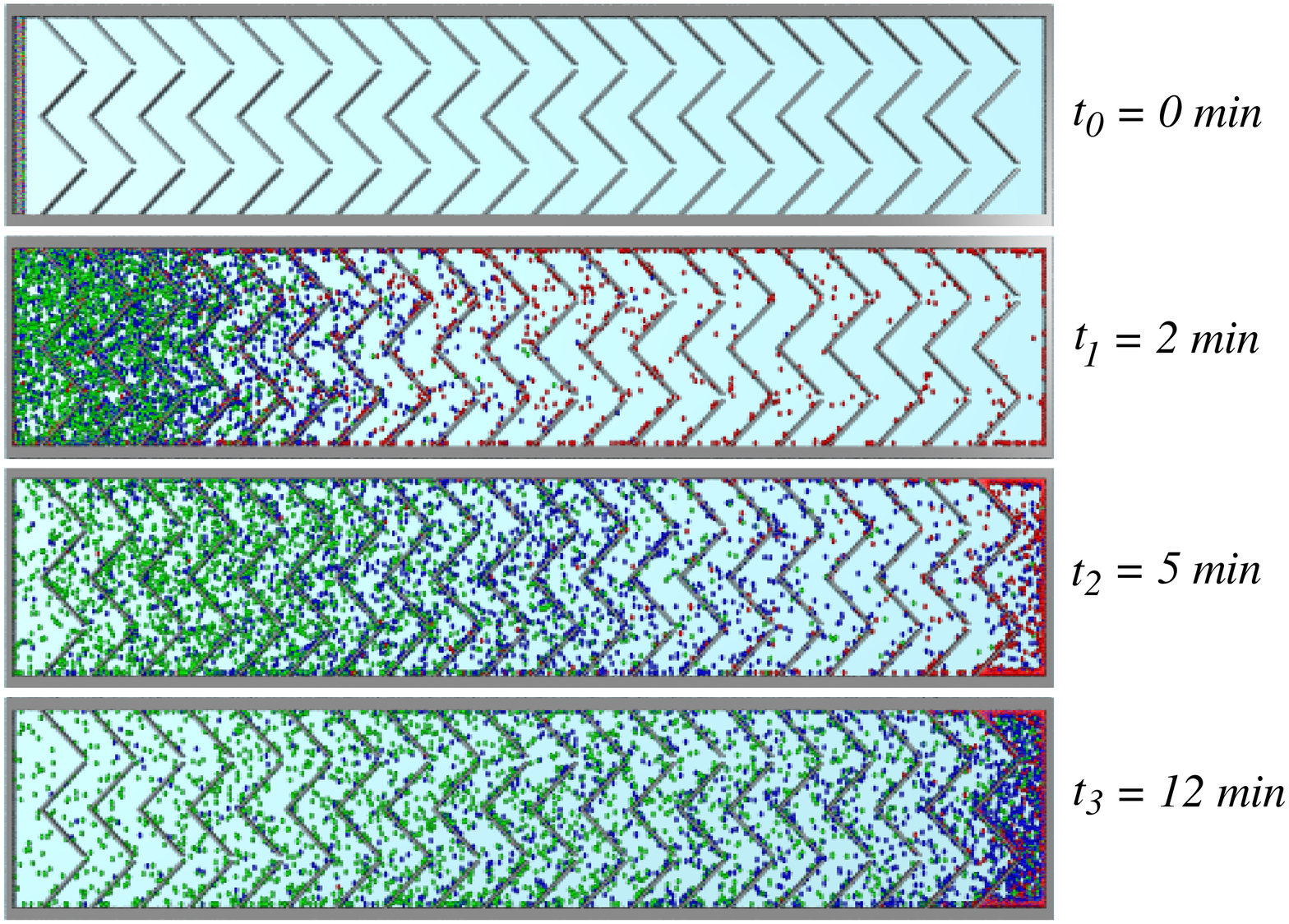}
\caption{(color online) Snapshots of simulated bacteria distribution vs time for a mixture including
 three {\it E. coli} mutants: (a) CheC497, $s_1$, represented by a red (black) symbol. (b) AW405 wild-type,
 $s_3$, blue (dark-grey) and (c) Unc602, $s_4$, green (light-grey), being the slowest. 
}  
\label{fig:snap}
\end{center}
\end{figure}   
\begin{figure}
\begin{center}
\includegraphics[angle=0, width=8.8cm]{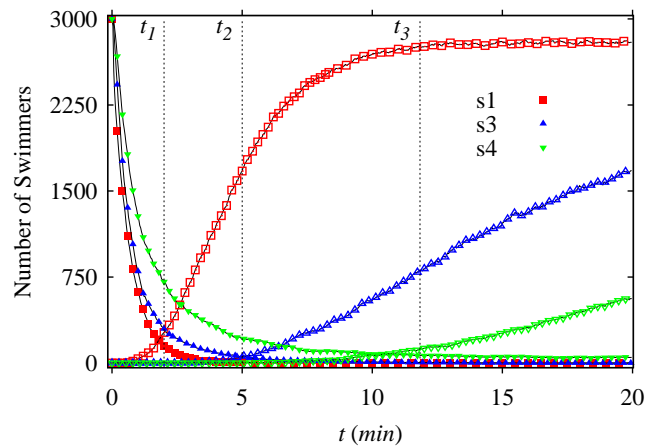}
\caption{(color online) Concentration of swimmers in the first/last chambers (decreasing and increasing curves, respectively)
 as a function of time, for three {\it E. coli} strains
  $s_1$, $s_3$ and $s_4$.  A  box $ L_x$ = 630 $\mu m$, $L_y$ = 64 $\mu m$ was divided into equal
 chambers by 20 columns of funnels, with 2 funnels each, of $l_f$ = 30 $\mu m$, $l_g$ = 2 $\mu m$ and $\theta$ = $60^\circ$. } 
\label{fig:sep}
\end{center}
\end{figure}
\begin{figure}
\begin{center}
\includegraphics[angle=0, width=8.8cm]{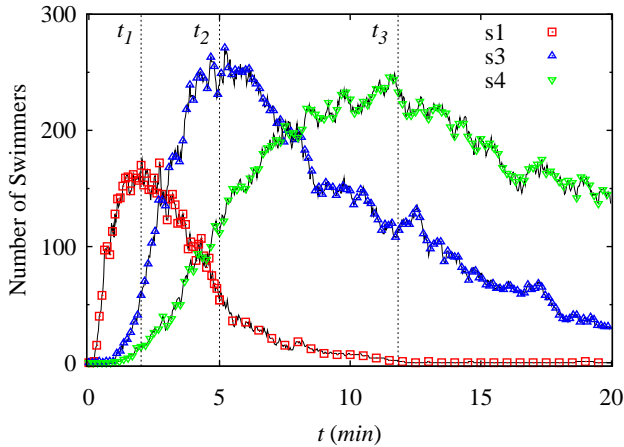}
\caption{(color online) Populations of swimmers $s_1$, $s_3$ and $s_4$ passing through the middle chamber, chamber number 11,
 vs  time.  All geometrical parameters are the same as in Fig.~\ref{fig:sep}.} 
\label{fig:sep_m}
\end{center}
\end{figure}

  If the three different strains of {\it E. coli} are inoculated at the same time, only part of the $s_4$ population
 remains in the first chamber 5 minutes after the inoculation, while none of the other two strains can be found there.
 See these details in Fig.~\ref{fig:sep}, where the numbers of swimmers at the first and last chambers are plotted vs time.
 The times chosen for the  snapshots in Fig.~\ref{fig:snap} are represented here by  vertical dotted lines. The three
 decreasing curves in Fig.~\ref{fig:sep}
  represent the time dependence of the various populations
   in the first chamber and the three increasing  curves  are the results measured in the last chamber. In chamber 21, 
 mutants of the strain $s_1$ begin to arrive just a minute after inoculation, about
$4$ minutes before  the arrival of the $s_3$ strain, opening a window of time large enough as to permit the  extraction
 of purified $s_1$ swimmers. Notice in Fig.~\ref{fig:snap}  that during the first two minutes $s_1$ could
 be extracted alone, in pure concentration, from the last five chambers, offering also a  wide spatial window.
     The same behavior can be noticed in  Fig~\ref{fig:sep_m} which shows the number of swimmers of each kind
 in the middle of the box  (chamber number 11) as a function of time. The strain $s_1$ has its highest concentration
 there  2 minutes after inoculation, whereas $s_3$ needs more than twice that time to get to the same place. Around
 5 minutes after inoculation, we could see that all chambers  present at least a small percentage of $s_4$. Then this
 kind of  chamber by chamber analysis, with our more accurate model, is helpful to design optimal separators and
 density-controlled mixers according to the swimming strategy and the required  percentage of purity.  Properly
 adjusting the geometry of the confinement box and the  separation of the funnels walls,  it is possible to obtain
  results in accordance with specified application requirements.

\section{Summary}

Arguably among the highest accomplishments in the  separation of motile cells are the sperm sorting achieved in
 a micro-fluidic system in 2003 by Cho and coworkers~\citep{Cho2003} and the sorting of {\it E. coli} cells by
 age/length reported by Hulme and collaborators in 2008~\citep{Hulme2008}. There are many other very recent achievements
 in sorting different (not self-propelled) particle  species  using electrokinetic effects or magnetic fields (Ref.~\citep{Bogunovic2012}
 and references therein). Motivated by this growing technological interest in finding efficient mechanisms able to separate mixtures
 of cell populations,   we have  studied numerically both the role of the free-space microorganism dynamics and the importance of
 the particular architecture of the  asymmetric obstacles.

We have considered the case of swimming bacteria, looking at the most relevant parameters characterizing the interaction between
 the microorganisms and their confining walls and searching for their optimal values. We have extended previous investigations
 by including such so far neglected experimental properties as the {\it distribution of run lengths},  the {\it rotational diffusion},
 and the {\it directional persistence}. Starting by examining the velocity correlation and the mean square displacement of unconfined
 self-propelled swimmers, we have shown how these properties are strongly dependent on run length and persistence angle. A smaller
 persistence angle implies a stronger memory of the initial direction and increases both the velocity correlation and the diffusion
 coefficient. Its effect is therefore similar to that of a longer run length. In the case of short runs, the diffusion coefficient
 is approximately proportional to the run duration whereas for the long-run swimmers, the effective diffusivity is controlled by
 the rotational diffusion, $D_r$. Consequently, the effective translational diffusion coefficient, $D$, turns out to be much smaller
 than what would result from a simple rescaling of the run length.

When the dynamics of free swimmers is incorporated into a spatially constrained environment, new effects emerge. Long run lengths
 and small tumble emergence angles lead to high wall accumulation and, consequently, to fast net displacement in the ratchet
 direction, which is clearly confirmed by the numerical calculations. In general, long permanence  near the walls and suitable
 wall architecture favor rectification.

It is important to point out that these results are rather robust and should remain valid for organisms such as  sperm cells
 (50 $\mu$m), algae (10 $\mu$m), bacteria (1 $\mu$m) or even viruses (100 nm) shaken by a drive of zero mean, such as pressure
 oscillation~\citep{Matthias2003}. In addition, the concept of mechanical reorientation of swimmer displacement via suitable
 wall architecture would allow one to envisage the design of surfaces with -phobic or -philic properties for adsorption or
 repulsion of microorganisms, respectively.

It would be also of high interest to perform separation measurements on swimmer mixtures with more complex motility patterns,
 such as the recently studied flicking marine bacteria {\it Vibrio alginolyticus}~\citep{Stocker2011}.
These measurements could also be performed during the growing stage of a heterogeneous wild-type {\it E. coli} population
 in order to select the spherical baby cells  by physical guidance with a  simple and easy-to-fabricate geometrical array  (Fig.~\ref{fig:snap}). 
 Such a process would eliminate the stress and damage associated with centrifugation techniques, a quality shared with the harder-to-build
 heart-shaped channel array proposed originally by 
Hulme and collaborators~\citep{Hulme2008}. 
This efficient separation by shape during the cell cycle is very important to study the cell motility properties by age both in free swimming and close to the walls of confinement (see Ref.~\citep{Gaston2011}).   

In order to guarantee the applicability of our method, we have to ensure that the smallest constriction remains larger than the smaller dimension of the swimmer.  Recently in  Ref.~\citep{Mannik2009, Guo2011}, the authors addressed the extreme case of growth and motion of bacteria in ultra-narrow constrictions with a size even smaller than their diameter. In this case, free-space swimmer strategy is clearly of no relevance although the architecture of the constraints and the particular form of the asymmetric walls can be of paramount importance.

We acknowledge financial support from CONICET, MINCyT,  SeCyT-UNC, and  FWO-MINCyT and KULeuven-UNC   bilateral projects and the unconditional support of F.A. Tamarit on this interdisciplinary project at Famaf, UNC. This work was partially supported by the Fonds de la Recherche Scientifique-FNRS, the Methusalem Funding of the Flemish Government, the Fund for Scientific Research-Flanders (FWO-Vlaanderen).

\bibliographystyle{biophysj}

\end{document}